\documentclass{jfm}
\usepackage{natbib,amsmath,amssymb,graphicx}
\usepackage[usenames,dvipsnames,svgnames,table]{xcolor}

\ifCUPmtlplainloaded \else
  \checkfont{eurm10}
  \iffontfound
    \IfFileExists{upmath.sty}
      {\typeout{^^JFound AMS Euler Roman fonts on the system,
                   using the 'upmath' package.^^J}%
       \usepackage{upmath}}
      {\typeout{^^JFound AMS Euler Roman fonts on the system, but you
                   dont seem to have the}%
       \typeout{'upmath' package installed. JFM.cls can take advantage
                 of these fonts,^^Jif you use 'upmath' package.^^J}%
      }
  \else
  \fi
\fi

\ifCUPmtlplainloaded \else
  \checkfont{msam10}
  \iffontfound
    \IfFileExists{amssymb.sty}
      {\typeout{^^JFound AMS Symbol fonts on the system, using the
                'amssymb' package.^^J}%
       \usepackage{amssymb}%
         
         \let\geq=\geqslant
      }{}
  \fi
\fi

\ifCUPmtlplainloaded \else
  \IfFileExists{amsbsy.sty}
    {\typeout{^^JFound the 'amsbsy' package on the system, using it.^^J}%
     \usepackage{amsbsy}}
    {\providecommand\boldsymbol[1]{\mbox{\boldmath $##1$}}}
\fi
\usepackage{overpic}
\usepackage{dcolumn}
\usepackage{epstopdf}
\usepackage{siunitx}
\usepackage{mathrsfs,amssymb}
\makeatletter
\def\amsbb{\use@mathgroup \M@U \symAMSb}
\makeatother
\expandafter\let\csname equation*\endcsname\relax
\expandafter\let\csname endequation*\endcsname\relax
\usepackage{amsmath}

\newcommand{\ve}[1]{\boldsymbol{#1}}
\renewcommand{\vec}[1]{\boldsymbol{#1}}

\newcommand{\uc}{u_{\rm c}}
\newcommand{\tauK}{\tau_{\rm K}}

\usepackage[bbgreekl]{mathbbol}

\begin{document}
\title{Inertial torque on a squirmer}
\author{F.~Candelier$^1$, J. Qiu$^2$, L. Zhao$^2$, G. Voth$^3$, B. Mehlig$^4$}
\affiliation{$^1$Aix Marseille Univ, CNRS, IUSTI, Marseille, France\\
	$^2$AML, Department of Engineering Mechanics, Tsinghua University, 100084 Beijing, China\\
	$^3$Department of Physics, Wesleyan University, Middletown, CT 06459, USA\\
	$^4$Department of Physics, Gothenburg University, 41296 Gothenburg, Sweden }

\date{}\maketitle
\begin{abstract}
A small spheroid settling in a quiescent fluid experiences an inertial torque that aligns it so that it settles with its broad side first. Here we show that an active particle experiences such a torque too, as it settles in a fluid at rest. For a spherical squirmer, the torque is  $\ve T^\prime = -{\tfrac{9}{8}} m_f (\ve v_s^{(0)} \wedge \ve v_g^{(0)})$ where $\ve v_s^{(0)}$ is the  swimming velocity, $\ve v_g^{(0)}$ is the settling velocity in the Stokes approximation, and $m_f$ is the equivalent  fluid mass. This torque aligns the swimming direction against gravity: swimming up is stable, swimming down is unstable. 
	\end{abstract}

\section{Introduction}

The motion of small plankton in the turbulent ocean is overdamped  \citep{visser2011small}. Accelerations play no role, and hydrodynamic forces and torques can be computed in the Stokes approximation. Turbulence rotates these small organisms, yet they manage to navigate upwards towards the ocean surface. Gyrotactic organisms make use of gravity to achieve this. These bottom-heavy swimmers experience a gravity torque that tends to align against the direction of gravity, so that they swim upwards \citep{kessler1985hydrodynamic,durham2013turbulence,gustavsson2015preferential}. 
Also density or shape asymmetries give rise to torques in the Stokes approximation that can change the swimming direction \citep{roberts1970geotaxis,jonsson1989vertical,roberts2002gravitaxis,candelier2016settling,roy2019inertial}.

Larger organisms accelerate the surrounding fluid as they move, and
this changes the hydrodynamic force the swimmer experiences  \citep{wang2012inertial,khair2014expansions,chisholm2016squirmer,redaelli2021unsteady,redaelli2022hydrodynamic}. Three different mechanisms
cause such fluid-inertia effects, a non-zero slip velocity (Oseen problem with non-dimensional parameter Re$_p$, the particle Reynolds number),
velocity gradients of the disturbance flow (Saffman problem, shear Reynolds number Re$_s$), and unsteady fluid inertia (with parameter
${\rm Re}_p {\rm Sl}$, where ${\rm Sl}$ is the Strouhal number). 

Fluid inertia gives rise to hydrodynamic torques. For a passive spheroid in spatially inhomogeneous flow, there are ${\rm Re}_s$-corrections to Jeffery's torque \citep{subramanian2005inertial,einarsson2015rotation,rosen2015numerical}.
A passive spheroid settling in a quiescent fluid experiences an inertial torque, a Re$_p$-effect. This Khayat-Cox torque tends to align the particle so that it settles with its broad side down \citep{brenner1961oseen,cox1965steady,khayat1989inertia,klett1995orientation,dabade2015effects,menon2017theoretical,lopez2017inertial,kramel2017nonpsherical,gustavsson2019theory,jiang2021inertial,cabrera2022experimental}. For a passive sphere, spherical symmetry ensures that the
Khayat-Cox torque vanishes. 
\begin{figure}
	\centering
	\begin{overpic}[width=10cm,clip]{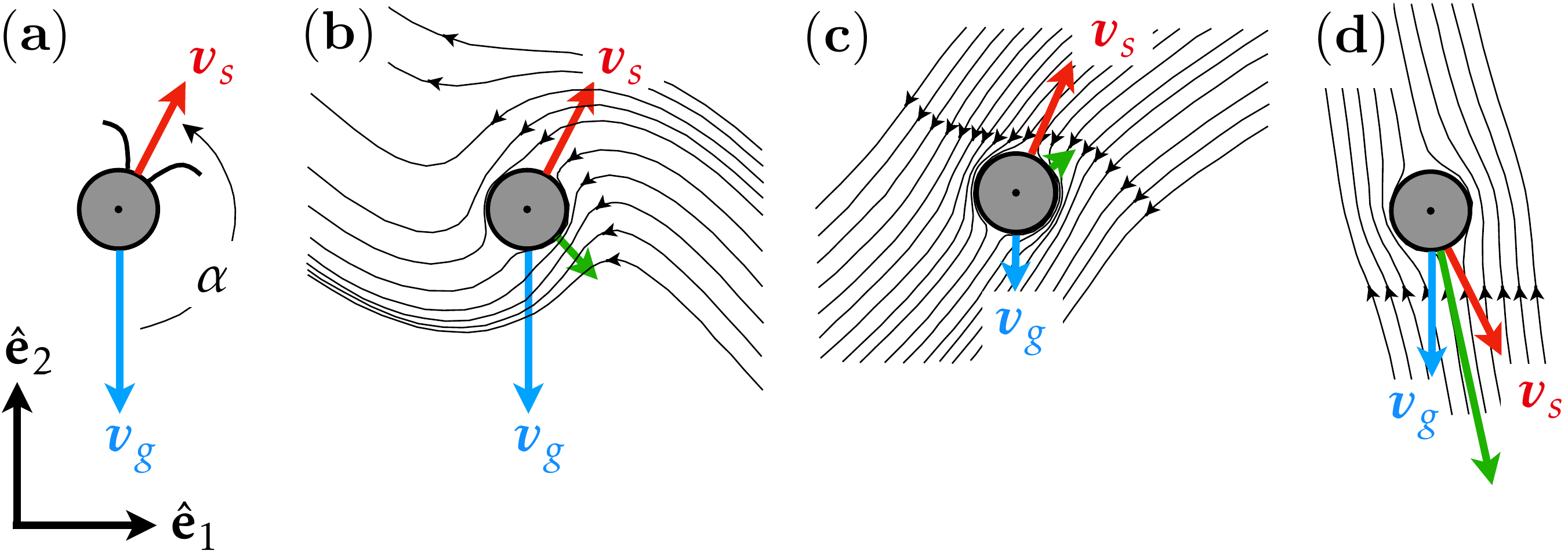}
	\end{overpic}
	\caption{\label{fig:schematic} ({\bf a}) Squirmer with swimming velocity $\ve v_s$ and settling velocity $\ve v_g$, see Section~\ref{sec:model}.   Gravity
	points in the negative $\hat{\bf e}_2$-direction. ({\bf b},{\bf c},{\bf d})~Disturbance flow created by a squirmer with $B_2=0$ (schematic). Shown are the flow lines in the frame that translates with the body.  The centre-of-mass velocity $\dot{\ve x} $ is shown in green. }
\end{figure}

In this paper, we show that a small spherical squirmer experiences an inertial torque analogous to the Khayat \& Cox torque when it settles in a quiescent fluid.  Using asymptotic matching, we calculate the torque to leading order in the particle Reynolds number
\begin{align}
	{\rm Re}_p = {a\uc}/{\nu}\,,
\end{align}
where $\uc$ is a  velocity scale, $a$ is the radius of the squirmer, and $\nu$ is the kinematic viscosity of the fluid. 
The calculation shows  that  the inertial torque does not
vanish for a spherical swimmer because  swimming breaks rotational symmetry. We describe how the torque aligns the squirmer, and compare its effect with gyrotactic torques, and with the Khayat-Cox torque for a nonspherical passive particle.

\section{Model}
\label{sec:model}
We consider a steady spherical squirmer, an idealised model for a motile microorganism developed by \cite{lighthill1952squirming} and \cite{blake1971spherical}. In this model,
one imposes
an active axisymmetric 
tangential surface-velocity field of the form
\begin{align}
\label{eq:vsurf}
	( B_1 \sin\theta + B_2 \sin\theta\cos\theta)\hat{\bf e}_\theta \,,
\end{align}
with parameters $B_1$ and $B_2$, and where $\theta$ is the angle between the swimming direction (unit vector $\ve n$) and the vector $\ve r$  from the particle centre to a point on its surface. The tangential unit vector at this point is denoted by $\hat{\bf e}_\theta$.
One distinguishes  two types of squirmers depending
on the parameter $\beta= B_2/B_1$ \citep{lauga2009hydrodynamics}: \lq  pushers\rq{}  ($\beta < 0$) and \lq pullers\rq{}  with $\beta > 0$.
 In the Stokes limit, a squirmer moving with velocity $\dot{\ve x}$ in a fluid at rest experiences the hydrodynamic force
\begin{align}
	{\vec{F}'}^{(0)}   = 6 \pi  \varrho_f \nu a \big( \tfrac{2}{3}B_1 \vec{n}-\dot{\ve x} \big)\,.
\end{align}
Here the superscript denotes the Stokes approximation, 
and $\varrho_f$
is the mass density of the fluid. Following \citet{candelier2019time}, we use a prime to indicate that this is the hydrodynamic force on the squirmer,
 due to the disturbance it creates.

Plankton tends to be slightly heavier than the fluid. Therefore we allow the squirmer to settle subject to the buoyancy force 
\begin{align}
\label{eq:F_g}
	\ve F_g = \tfrac{4\pi}{3}a^3 (\varrho_s-\varrho_f)\ve g\,,
\end{align}
where $\varrho_s$ is the mass density of the squirmer, and $\ve g$ is the gravitational acceleration. In the overdamped limit, the steady centre-of-mass velocity of the squirmer
is determined by the zero-force condition ${\vec{F}'}^{(0)} +\ve F_g=\ve 0$. This yields
$\dot{\ve x} = \tfrac{2}{3} B_1 \ve n + \tfrac{2}{9}\big(\tfrac{\varrho_s}{\varrho_f}-1\big) \tfrac{a^2}{\nu}\ve g\equiv \ve v_s^{(0)} + \ve v_g^{(0)}$.
Again, the superscript denotes the Stokes limit. In this limit, the squirmer experiences no torque in a fluid at rest, ${\ve T^{\prime}}^{(0)} = \ve 0$.

\section{Inertial torque}
\label{sec:torque}
Assume that the squirmer swims  with swimming velocity $\ve v_s$ and settles with settling velocity $\ve v_g$. The angle
between $\ve v_s$ and $\ve v_g$ is denoted by $\alpha$, as shown in Fig.~\ref{fig:schematic}({\bf a}). Symmetry dictates
the form of the inertial torque $\ve T^\prime$. It has  the units mass $\times$ velocity$^2$. Since the torque is an axial vector, it must be proportional
to the vector product between the two velocities. The torque can therefore be written as
\begin{equation}
\label{eq:torque3}
	{\ve T^\prime}^{(1)} = C m_f\,(\ve v_s \wedge \ve v_g)\,,
\end{equation}
where $m_f = \tfrac{4\pi}{3}a^3 \varrho_f$ is the equivalent fluid mass, $C$ is a non-dimensional constant, and the superscript indicates that this is the first  inertial correction to the torque. Eq.~(\ref{eq:torque3})~says that torque
vanishes when the swimmer swims against gravity [$\alpha = \pi$ in Fig.~\ref{fig:schematic}({\bf a})], and when it swims in the direction of gravity ($\alpha=0$). Bifurcation theory
implies that one of these fixed points is stable, the other one unstable. The sign of the coefficient $C$ determines which of the two is the stable fixed point.

Inertial torques can be understood as a consequence of advection of fluid momentum. In the frame translating with the squirmer, far-field momentum is advected by the transverse disturbance flow generated by the squirmer.  At non-zero ${\rm Re}_p$, the head of the squirmer -- the north pole of the axial velocity field (\ref{eq:vsurf}) -- experiences more drag than its rear, because some of the momentum imparted to the fluid by the head is advected to the trailing end, in the direction transverse to gravity. So when $\ve v_s$ is not co-linear with $\ve v_g$ there is an inertial torque which rotates the swimmer so that $\ve v_s$ becomes closer to anti-parallel with $\ve v_g$.   Comparing with Eq.~(\ref{eq:torque3}), this means that the coefficient $C$ must be negative. 
Note that the mechanism described above  is the same  that creates Khayat-Cox torques on non-spherical passive particles sedimenting in quiescent fluid.  For a fibre, for example, the  far-field momentum is advected by the transverse flow along the fibre, leading to a torque that aligns the fibre perpendicular to gravity \citep{khayat1989inertia}.

\section{Perturbation theory for the coefficient $C$}
The inertial torque is computed from
\begin{align}
\label{eq:torque}
	{{\ve  T}^\prime}^{(1)}=  \int_{\mathscr{S}}\!\! \ve r\wedge(\bbsigma^{(1)}{\rm d}\ve s) \,,
\end{align}
where  $\sigma_{mn}^{(1)}  = -p^{(1)}\delta_{mn} + 2 \mu S_{mn}^{(1)}$  are the elements of the stress tensor $\bbsigma^{(1)}$ with  pressure $p^{(1)}$,  $S_{mn}^{(1)}$ are the elements of the strain-rate tensor of the disturbance flow, and $\mu=\varrho_f \nu$ is the dynamic viscosity. The integral goes over the particle surface $\mathscr{S}$,
$\ve r$ is the vector from the particle centre to a point on the particle surface, and
${\rm d}\ve s$ is the outward surface normal at this point. 
In the Stokes approximation the torque vanishes, ${{\ve T}^\prime}^{(0)}=\ve 0$, as mentioned above.

The disturbance stress tensor is determined by solving the steady Navier-Stokes equations for the incompressible disturbance flow $\ve w$,
\begin{equation}\label{eq:eom}
		-{\rm Re}_{p} \dot{\ve x} \cdot \boldsymbol{\nabla} \ve{w}+{\rm Re}_{\rm{p}} \: \ve{w} \cdot \boldsymbol{\nabla} \ve{w} = - \boldsymbol{\nabla} p + \boldsymbol{\triangle} \ve{w} \,,
	\end{equation}
	with boundary conditions
	$\ve{w} = \dot{\ve x}  + ( B_1 \sin\theta + B_2 \sin\theta\cos\theta)\hat{\bf e}_\theta$ for $|\ve r|=1$, and {$\ve{w} \to \ve{0}$} as  $|\ve r|\to \infty$.
Here we assumed that the squirmer has no angular velocity. We non-dimensionalised Eq.~(\ref{eq:eom}) using the radius $a$ of the squirmer as a length scale, and
with the velocity scale $\uc=v_g^{(0)}$. Forces are non-dimensionalised by $\mu a \uc$, and torques by $\mu a^2 \uc$. The acceleration terms on the l.h.s. of Eq.~(\ref{eq:eom}) are  singular perturbations of the r.h.s., the Stokes part \citep{hinch1995perturbation}. 
We use matched asymptotic expansions in ${\rm Re}_p$ to determine the solution for small Re$_p$. Near the squirmer, one expands
\begin{equation}
	{\ve{w}}_{\mbox{\scriptsize in}} = {\ve{w}}_{\mbox{\scriptsize in}}^{(0)} + {\rm Re}_p  \: {\ve{w}}_{\mbox{\scriptsize in}}^{(1)} + \ldots
	\quad \mbox{and} \quad  
	{p}_{\mbox{\scriptsize in}} = {p}_{\mbox{\scriptsize in}}^{(0)} + {\rm Re}_p  \: {p}_{\mbox{\scriptsize in}}^{(1)} +   \ldots \:.
	\label{eq:flow_plotted}
\end{equation}
This inner expansion is matched, term by term, to an outer expansion 
\begin{equation}
	\hat{{\ve{w}}}_{\mbox{\scriptsize out}}  = \hat{\mathcal{T}}_{\mbox{\scriptsize reg}}^{(0)}  + {\rm Re}_p( \hat{\mathcal{T}}_{\mbox{\scriptsize reg}}^{(1)} + \hat{\mathcal{T}}_{\mbox{\scriptsize sing}}^{(1)} )+ \ldots \:.
	\label{W_outer1}
\end{equation}
Here $\hat{\mathcal{T}}_{\mbox{\scriptsize reg}}^{(0,1)}$ are regular terms in the outer expansion, while 
$\hat{\mathcal{T}}_{\mbox{\scriptsize sing}}^{(1)}$ is  singular in $\ve k$-space, proportional to $\delta(\ve k)$ \citep{meibohm2016angular}.
The outer solution is obtained by replacing the boundary condition on the surface of the squirmer by a singular source term in Eq.~(\ref{eq:eom}), a  Dirac $\delta$-function \citep{schwartz1966theorie}
with amplitude $\ve F^{(0)} = -6\pi(\tfrac{2}{3} B_1 \ve n-\dot{\ve x} )$.  Since the non-linear term (quadratic in $\ve w$) is negligible far from the particle, the resulting equation can be solved by Fourier transform, yielding 
explicit expressions for $\hat{\mathcal{T}}_{\mbox{\scriptsize reg}}^{(0,1)}$ and  $\hat{\mathcal{T}}_{\mbox{\scriptsize sing}}^{(1)}$ which serve as boundary conditions
for the inner problems.

The inner problem to order ${\rm Re}_p^0$ is the homogeneous Stokes problem
\begin{subequations}
	\label{eq:eps0}
	\begin{align}
		- \boldsymbol{\nabla} {p}_{\mbox{\scriptsize in}}^{(0)} + \boldsymbol{\triangle}  
		{\vec{w}}_{\mbox{\scriptsize in}}^{(0)}  =  \vec{0} \:,\:\: \boldsymbol{\nabla} \cdot  {\vec{w}}_{\mbox{\scriptsize in}}^{(0)}= \vec{0}\,,
		\label{w0_0}
	\end{align}
  with boundary conditions
	\begin{align}
		&{\vec{w}}_{\mbox{\scriptsize in}}^{(0)} =   \dot{\ve x} \! +\! ( B_1 \sin\theta \!+ \!B_2 \sin\theta\cos\theta)\hat{\bf e}_\theta  \: \:\mbox{for}\:\: { |\vec r|=1}\,,
		 \:\: {\vec{w}}_{\mbox{\scriptsize in}}^{(0)} \sim  \vec{\mathcal{T}}_\mathrm{reg}^{(0)} \: \:\mbox{as}\:\:  |\vec r| \to \infty\:.
		\label{w_0_BC}
	\end{align}
\end{subequations}
This problem is solved in the standard fashion using Lamb's solution \citep{happel1981low}.
The ${\rm Re}_p^1$-order inner problem is inhomogeneous:
\begin{subequations}
	\label{eq:eps1}
	\begin{align}
		&- \boldsymbol{\nabla} {p}_{\mbox{\scriptsize in}}^{(1)} + \boldsymbol{\triangle}  
		{\vec{w}}_{\mbox{\scriptsize in}}^{(1)}  = - {\rm Re}_p \dot{\ve x} \cdot \boldsymbol{\nabla} {\vec{w}}_{\mbox{\scriptsize in}}^{(0)} +  {\rm Re}_p {\vec{w}}_{\mbox{\scriptsize in}}^{(0)}\cdot \boldsymbol{\nabla} {\vec{w}}_{\mbox{\scriptsize in}}^{(0)}\,,\quad 	\boldsymbol{\nabla} \cdot  {\vec{w}}_{\mbox{\scriptsize in}}^{(1)}=  0\,,
		\label{eq_order1}
		\\&
		{\vec{w}}_{\mbox{\scriptsize in}}^{(1)} =   \vec{0}    \: \:\mbox{for}\:\: { |\vec r|=1}
		\quad \mbox{and} \quad 
		\quad {\vec{w}}_{\mbox{\scriptsize in}}^{(1)} \sim  \vec{\mathcal{T}}_\mathrm{reg}^{(1)} +\vec{\mathcal{T}}_\mathrm{sing}^{(1)}  \:\:\mbox{for}\:\:  |\vec r|\to \infty\:. 
		\label{eq_order_1_BC}
	\end{align}
\end{subequations}
To solve Eqs.~(\ref{eq:eps1}), we make the ansatz ${\vec{w}}_{\mbox{\scriptsize in}}^{(1)} =( \vec{w}_{\mbox{\scriptsize p}}
+  \vec{\mathcal{T}}_\mathrm{sing}^{(1)})+ \vec{w}_{\mbox{\scriptsize h}}$, where $ \vec{w}_{\mbox{\scriptsize p}}$ is a particular solution, and $ \vec{w}_{\mbox{\scriptsize h}}$ is the homogeneous solution of Eqs.~(\ref{eq:eps1}). For the pressure we write $p_{\rm in}^{(1)} = p^{(1)}_{\rm p}+p^{(1)}_{\rm h}$.
We first determine the particular solution $ \vec{w}_{\mbox{\scriptsize p}}$ and $p^{(1)}_{\rm p}$ using Fourier transform. 
Then $\vec{w}_{\mbox{\scriptsize h}}$ and $p^{(1)}_{\rm h}$ are determined using Lamb's solution. The boundary condition
for $ \vec{w}_{\mbox{\scriptsize h}}$ is $ \vec{w}_{\mbox{\scriptsize h}}=- \vec{w}_{\mbox{\scriptsize p}}-  \vec{\mathcal{T}}_\mathrm{sing}^{(1)}$ on the particle surface. 
Having obtained ${\vec{w}}_{\mbox{\scriptsize in}}^{(1)}$, we compute the torque from Eq.~(\ref{eq:torque}). The torque comes from the particular solution of the 
first-order inner problem. For a passive spherical particle, spherical symmetry ensures that the particular solution does not contribute to the torque. Swimming breaks
spherical symmetry, and this is the reason that torque does not vanish.
 Given $p^{(1)}$  and $\ve w_{\rm in}^{(1)}$, we can determine
the inertial correction to the stress tensor, $\bbsigma^{(1)}$. Integrating it over the particle surface as specified in Eq.~(\ref{eq:torque}), we find to leading order in ${\rm Re}_p$
\begin{align}
\label{eq:torque_result}
	{\ve T^\prime}^{(1)} = -\tfrac{3\pi}{2} {\rm Re}_p (\ve v_s^{(0)} \wedge \ve v_g^{(0)})\,.
	\end{align}
In dimensional units, this corresponds to ${\ve T^\prime}^{(1)} = -{\tfrac{9}{8}} m_f (\ve v_s^{(0)} \wedge \ve v_g^{(0)})$. 
The coefficient $C=-{\tfrac{9}{8}}$ is negative, as predicted by the argument summarised in Section \ref{sec:torque}. So a spherical organism swimming downwards experiences a torque that tends to turn it upwards, causing the organism to swim against gravity.

\section{Direct numerical simulations}
\label{sec:dns}
\begin{figure}
	\centering
	\begin{overpic}[width=12cm,clip]{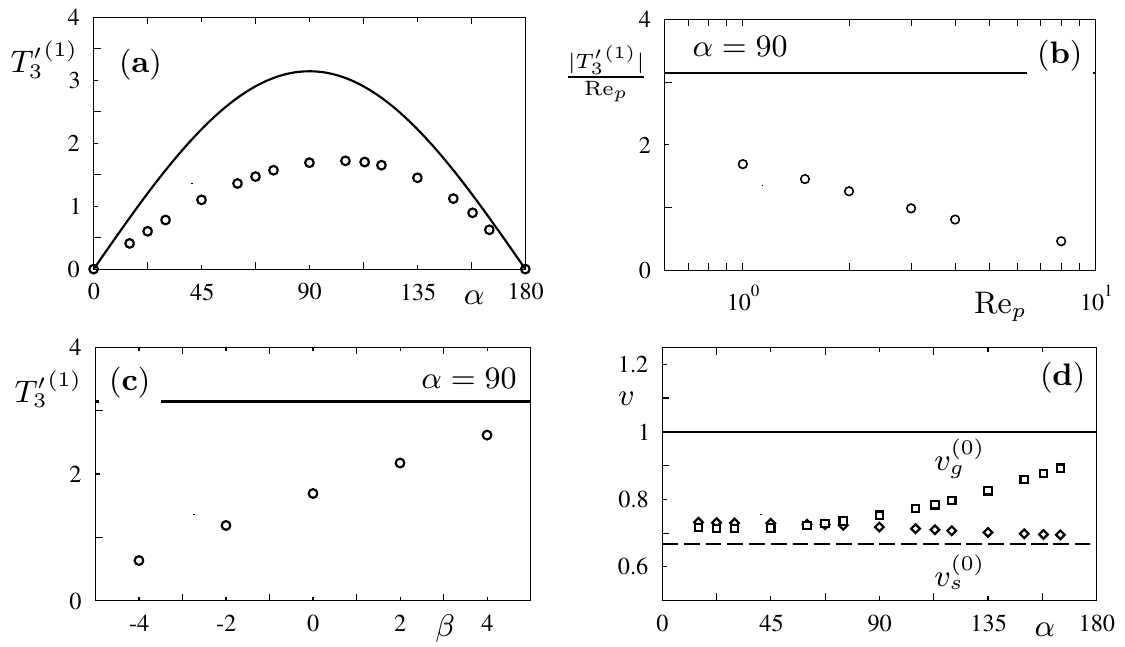}
			\end{overpic}
		\caption{\label{fig:torque}({\bf a})  Non-dimensional inertial torque ${\ve T^\prime}^{(1)}={T^\prime_3}^{(1)} \hat{\bf e}_3$ on a spherical squirmer. Shown is the theory, Eq.~(\ref{eq:torque_result})  (solid line), in comparison with direct numerical-simulation results (Section~\ref{sec:dns},~$\circ$).
		The torque was non-dimensionalised with the factor $\mu a^2 \uc$, where $\uc=v_g^{(0)}$. The angle $\alpha$ is defined in Fig.~\ref{fig:schematic}({\bf a}).
		Parameters:  $B_1=1$, $B_2=0$, $v_s^{(0)}= 2/3$, $v_g^{(0)} =1$, and Re$_p=1$.  ({\bf b})~Non-dimensional torque 
		for $\alpha=90$  as a function
		of ${\rm Re}_p$, other parameters same as in panel~({\bf a}). Also shown is the theory, Eq.~(\ref{eq:torque_result}).
		 ({\bf c})~Non-dimensional torque 
		for $\alpha=90$ as a function of $\beta=B_2/B_1$, other parameters the same as in panel~({\bf a}). ({\bf d})~Non-dimensional swimming speed ($\diamond$) and settling speed ($\Box$) from direct numerical simulations for 
		the same parameters as in panel~({\bf a}). Also shown are the Stokes estimates for $v_s^{(0)}$ (dashed line) and $v_g^{(0)}$  (solid line).
		}
\end{figure}

We solved the three-dimensional Navier-Stokes equations for the incompressible flow using
an immersed-boundary method  \citep{peskin2002immersed}. The interaction between squirmer and fluid was implemented by the direct-force method \citep{uhlmann2005immersed}: in order to satisfy the boundary condition (\ref{eq:vsurf}), the algorithm calculates the predicted fluid velocity on the surface of the squirmer. Based on the mismatch between the predicted velocity and Eq.~(\ref{eq:vsurf}), an appropriate immersed-boundary force is  applied to the fluid phase to maintain the boundary conditions (\ref{eq:vsurf}) on the surface of the squirmer. 
We implemented the improved algorithm described in \citep{kempe2012improved,breugem2012second,lambert2013active}, because it is more precise for 
nearly-neutrally buoyant particles.  We used a cubic computational domain of side length $L= 20a$ with periodic boundary conditions. This is large enough to account for convective inertia for Re$_p>1$. The computational domain was discretised using a cubic mesh with resolution $\Delta x$. The Navier-Stokes equations were integrated using a second-order Crank-Nicholson scheme for the time-integration \citep{kim2002implicit} with time step $\Delta t$, while the motion of the squirmer was integrated using a second-order Adams-Bashforth method \citep{hairer2000solving}. 

The numerical simulation of solid-body motion in a fluid  is challenging at small Re$_p$. 
The mesh resolution $\Delta x$ must be fine enough to resolve the shape of the body, so that the viscous stresses near its  surface are accurately represented \citep{andersson2019forces}. In addition, the time step $\Delta t$ must be small enough to resolve the viscous diffusion of the disturbance, $\Delta t < \Delta x^2/\nu$.
In Appendix \ref{app:A} we briefly describe our convergence checks. We found that our algorithm fails
to converge for Re$_p$ smaller than unity. In the following we discuss our numerical results for Re$_p\geq 1$.
  
 To determine the torque, we froze the orientation of the squirmer at a given angle~$\alpha$, but allowed the squirmer to translate.
It was initially at rest. We measured  the centre-of-mass velocity and the torque after the transient, when
 the disturbance flow was fully established.
  Figure~\ref{fig:torque}({\bf a})  shows the numerical results for the inertial torque on a spherical squirmer for Re$_p=1$, in comparison with the theory (\ref{eq:torque_result}).
The remaining parameter values used in the simulations are quoted in the Figure caption. 
Although the theory is valid for Re$_p \ll 1$, it nevertheless  agrees qualitatively with the numerical results for Re$_p=1$. This is encouraging, because
it allows as to draw qualitative conclusions about the effects of the torque on small plankton (Section \ref{sec:conc}). 
 Panel ({\bf a}) shows that the theory overestimates the amplitude of the torque by a factor of two, but that its angular dependence is roughly the same. 
 We note, however, that the numerical results exhibit an asymmetry in their dependence on $\alpha$. Since the small-Re$_p$ theory yields a symmetric angular dependence of the torque, we attribute the asymmetry to higher-order Re$_p$-corrections. Panel ({\bf b}) confirms that the difference between theory and numerical simulation increases for larger Re$_p$, as expected. 
 
The small-Re$_p$ theory (\ref{eq:torque_result}) says that the torque is independent of $\beta$. Panel ({\bf c}) shows that this is not the case for the numerical results at Re$_p=1$.
This suggests, again, that higher-Re$_p$ corrections  matter at Re$_p=1$, and that they exhibit a $\beta$-dependence. 
Another indication that higher-order ${\rm Re}_p$-corrections may be important comes from measuring settling and swimming speeds in the numerical simulations.
We extracted the swimming speed using $\dot{\ve x} = v_s \ve n - v_g \hat{\bf e}_2$. Solving for $v_s$ gives
$v_s = \dot{\ve x}\cdot \hat{\bf e}_1/(\ve n\cdot \hat{\bf e}_1)$. 
Fig.~\ref{fig:torque}({\bf d}) compares the measured swimming and settling speeds.
The settling speed is substantially smaller than the Stokes estimate, consistent with a significant ${\rm Re}_p$-correction.
The swimming speed is much closer to the Stokes estimate. This is because the data shown is for $\beta=0$, and the known ${\rm Re}_p$-corrections to
the swimming speed~\citep{khair2014expansions},
\begin{equation}
\label{eq:vs_khair}
\ve v_s = \tfrac{2}{3} B_1 \ve n[1-\tfrac{3\beta}{20} {\rm Re}_p +(\tfrac{\beta}{8}+\tfrac{11987}{470400}\beta^2) {\rm Re}_p^2+\ldots ]\,,
\end{equation}
vanish for $\beta=0$.

\section{Conclusions}
\label{sec:conc}
We showed that a spherical squirmer settling in a fluid at rest experiences an inertial torque, and computed the torque using matched asymptotic expansions. The calculation is similar to that of \cite{cox1965steady} for the inertial torque on a  nearly spherical, passive particle settling in a quiescent fluid. This  torque vanishes for a passive sphere, a consequence of spherical symmetry.   A spherical swimmer experiences an inertial torque because swimming
breaks this symmetry. The torque causes the squirmer to align with gravity so that it swims upwards. In other words,
this torque acts just like Kessler's gyrotactic torque for bottom-heavy organisms.

For plankton, the effect of the inertial torque is much smaller than the gyrotactic torque, at least for spherical shapes. We can see this by comparing
the corresponding reorientation times. This time scale is defined as $\tau_I = \tfrac{1}{2} (8\pi \mu a^3)/T_{\rm max}$, where $8\pi \mu a^3$ 
is the rotational resistance coefficient for a sphere  \citep{kim2013microhydrodynamics}, and $T_{\rm max}$ is the maximal magnitude of the torque.
For the inertial torque, one obtains  $ \tau_I ={8} \nu/({3}v_s^{(0)} v_g^{(0)})$ (this and all following expressions are quoted in dimensional units).
The reorientation time for the gyrotactic torque is  $\tau_G = 3\varrho_s\nu/(\varrho_fgh)$ \citep{pedley1987orientation}, where $h$ is the offset between the centre-of-mass and 
the geometrical centre of the squirmer, and $g = |\ve g|$.
The ratio of these time scales is  $\tau_I / \tau_G \sim gh/v_s^{(0)}v_g^{(0)}$, assuming $\varrho_s \approx \varrho_f$. Taking
$h\sim 10^{-7}$~m [Table~1 in~\citep{kessler1986individual}], we see that swimming and settling speeds need to be of the order mm/s for the reorientation
times to be comparable.
For small plankton, typical speeds tend to be much smaller \citep{kessler1986individual}. 
For larger organisms, however,  the inertial torque can be significant. With  typical values for
a small copepod \citep{titelman2003motility}, $v_s = 1\, \rm{mm/s}$, $v_g = 0.2\, \rm{mm/s}$, as well as $\nu=10^{-6}$m$^2$/s,  one finds an inertial reorientation time of the order of $\tau_I \sim 10\,$s. Kolmogorov times for ocean turbulence range from $\tauK=\sqrt{\nu/\mathscr{E}}=100$\,s for dissipation rate per unit mass $\mathscr{E}=10^{-6}$\,cm$^2$/s$^3$
to $\tauK=1$\,s for $\mathscr{E}=10^{-2}$\,cm$^2$/s$^3$. So the non-dimensional reorientation parameter $\Psi =\tau_I/\tauK$~\citep{durham2013turbulence} ranges from $0.1$ for weak turbulence to $10$ for strong turbulence.  
The Reynolds number is of order Re$_p \sim 1$ for speeds of the order of $1$\,mm. This means that the Re$_p$-perturbation theory does not strictly apply, but we can nevertheless
conclude that for weak turbulence, the inertial torque can have a significant effect on the angular dynamics of the organism. 

Some motile microorganisms are non-spherical \citep{berland1995observations,faust2002identifying,smayda2010adaptations}. It has been
suggested that this can give additional contributions to the torque \citep{qiu2022gyrotactic}. Since the boundary conditions are different, and since swimming breaks fore-aft symmetry, these additional contributions may be different from the Khayat-Cox torque for passive particles. However, we expect that the torque is still determined by the same
physical mechanism, advection of fluid-momentum transverse to gravity.  This may give rise to terms proportional to $\sin(2\alpha)$, whereas the torque
is proportional to $\sin(\alpha$) for the spherical squirmer. 
To  make these speculations definite, one could compute the inertial torque for a nearly spherical squirmer in perturbation theory.
A second open question is to determine the inertial torque for bottom-heavy, non-spherical organisms, the analogue of the inertial torque on passive particles
with mass-density asymmetries \citep{roy2019inertial}. 

More generally, although the small-Re$_p$ perturbation theory may become
quantitatively inaccurate for Reynolds numbers of order unity -- where the torque begins to make a significant difference -- the results tell us which non-dimensional parameters matter, and how to reason about the effect of  boundary conditions, and the symmetries of the problem. The calculation illustrates
the conceptual insight that  the  inertial torque comes from fluid motion transverse to the direction of
gravity. Fluid momentum in this direction is advected along the swimmer by the transverse fluid velocity, resulting in a torque.
In our case, the boundary conditions are different from those for passive particle, and so is the symmetry of the problem, because swimming breaks fore-aft symmetry.
Nevertheless, the fundamental mechanism generating the torque is the same.

{\em Acknowledgements.} BM was supported by Vetenskapsr\aa{}det (grant no.  2021-4452) and by the Knut-and-Alice Wallenberg Foundation (grant no.  2019.0079). 
LZ was supported by the National Natural Science Foundation of China (grant nos. 11911530141 and 91752205). This research was also supported in part by  a collaboration grant from the joint China-Sweden mobility programme [National Natural Science Foundation of China (NSFC)-Swedish Foundation for International Cooperation in Research and Higher Education (STINT)],  grant nos. 11911530141 (NSFC) and CH2018-7737 (STINT).

\appendix
\section{Details regarding the direct numerical simulations}
\label{app:A}
\begin{figure}
\centering
	\begin{overpic}[width=9cm,clip]{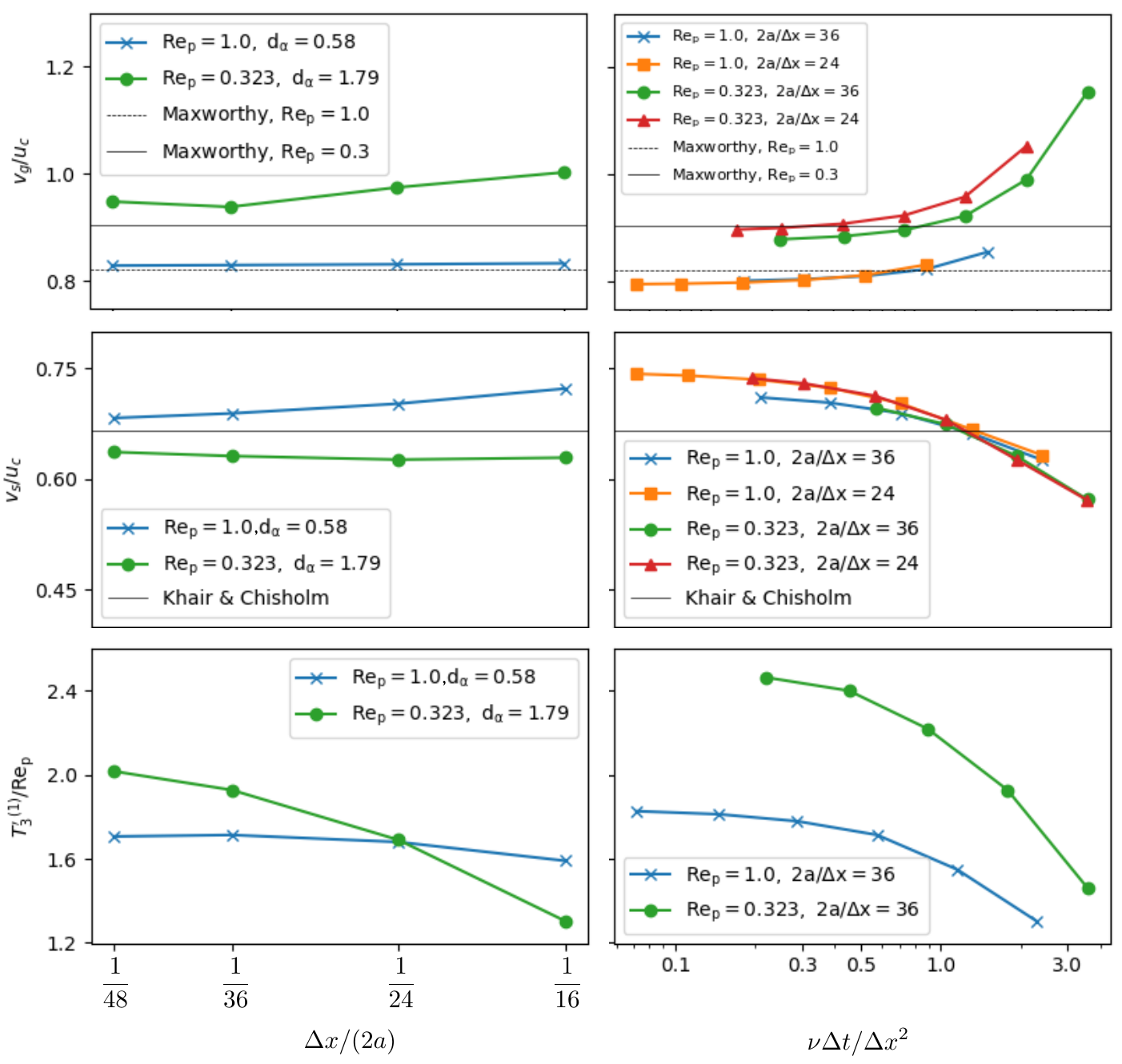}
	\end{overpic}
	\caption{\label{fig:convergence} Convergence tests changing mesh resolution $\Delta x$ (left column) and changing integration-time step $\Delta t$ (right column). Settling speed of a passive sphere (first row), compared with  the experimental data of Maxworthy, extracted from Fig.~4 of \citep{veysey2007simple}. Swimming speed of neutrally buoyant spherical squirmer with $\beta=0$, compared with Eq.~(\ref{eq:vs_khair}), \citep{khair2014expansions}, second row. The third row shows the inertial torque. }
\end{figure}

 To find out the required mesh and time resolution, we considered two test cases at $\rm{Re}_p = 1$: a passive sphere settling under gravity, and a neutrally buoyant 
 squirmer with $\beta = 0$. To check convergence as the mesh resolution increases, we changed  $2a/\Delta x$ ranging from 16 to 48,  keeping $\nu\Delta t/\Delta x^2= 0.58$ 
 constant. Settling and swimming speeds reached a plateau when we increased the mesh resolution $2a/\Delta x$. 
The settling speed varied about 0.1\% and swimming speed varied about 1\% when we changed $2a/\Delta x$ from 36 to 48. 

We then checked for convergence as the step size $\Delta t$ was reduced, for fixed $2a/\Delta x=36$. 
Again, both  settling and swimming speeds reached plateaus as   $\nu\Delta t/\Delta x^2$ decreased.
When we halved  $\nu\Delta t/\Delta x^2$ from 0.58 to 0.29, the settling and swimming speeds varied about 1\% and 2\%, respectively. Therefore, we used $2a/\Delta x=36$ and
 $\Delta t = 0.58\Delta x^2 / \nu$ for our numerical simulations at $\rm{Re}_p = 1.0$ that are shown in the main text. At these parameter values, the simulated settling speed of a passively settling particle is about 3\% larger than the measurement of Maxworthy, taken from
 Fig.~4 of \citep{veysey2007simple}.  The swimming speed of the squirmer is about half a percent larger than the theoretical value $v_s^{(0)} = 2B_1/3$.
  The convergence  for the inertial torque was slightly worse. The torque varied about 0.5\% when we changed $2a/\Delta x$ from 36 to 48 for $\nu\Delta t/\Delta x^2= 0.58$,  and it varied about  4\% when we  halved  $\nu\Delta t/\Delta x^2$ from 0.58 to 0.29 for $2a/\Delta x=36$.
  We also performed convergence checks for Re$_p=0.32$, reducing the particle Reynolds number by increasing $\nu$. We found that the values of $\Delta x$ and $\Delta t$ quoted above are too large for the numerical scheme to converge at this Reynolds number. Therefore we only show results for Re$_p\geq 1$ in the main text.


\begin{thebibliography}{53}
\expandafter\ifx\csname natexlab\endcsname\relax\def\natexlab#1{#1}\fi

\bibitem[Andersson \& Jiang(2019)]{andersson2019forces}
{\sc Andersson, Helge~I \& Jiang, Fengjian} 2019 Forces and torques on a
  prolate spheroid: Low-reynolds-number and attack angle effects. {\em Acta
  Mechanica\/} {\bf 230}~(2), 431--447.

\bibitem[Berland {\em et~al.\/}(1995)Berland, Maestrini \&
  Grzebyk]{berland1995observations}
{\sc Berland, Brigitte~R, Maestrini, Serge~Y \& Grzebyk, Daniel} 1995
  Observations on possible life cycle stages of the dinoflagellates dinophysis
  cf. acuminata, dinophysis acuta and dinophysis pavillardi. {\em Aquatic
  Microbial Ecology\/} {\bf 9}~(2), 183--189.

\bibitem[Blake(1971)]{blake1971spherical}
{\sc Blake, J.~R.} 1971 A spherical envelope approach to ciliary propulsion.
  {\em J. Fluid Mech.\/} {\bf 46}, 199--208.

\bibitem[Brenner(1961)]{brenner1961oseen}
{\sc Brenner, H.} 1961 The {O}seen resistance of a particle of arbitrary shape.
  {\em J. Fluid Mech.\/} {\bf 11}, 604--610.

\bibitem[Breugem(2012)]{breugem2012second}
{\sc Breugem, Wim-Paul} 2012 A second-order accurate immersed boundary method
  for fully resolved simulations of particle-laden flows. {\em Journal of
  Computational Physics\/} {\bf 231}~(13), 4469--4498.

\bibitem[Cabrera {\em et~al.\/}(2022)Cabrera, Sheikh, Mehlig, Plihon, Bourgoin,
  Pumir \& Naso]{cabrera2022experimental}
{\sc Cabrera, F., Sheikh, M.~Z., Mehlig, B., Plihon, N., Bourgoin, M., Pumir,
  A. \& Naso, A.} 2022 Experimental validation of fluid inertia models for a
  cylinder settling in a quiescent flow. {\em Phys. Rev. Fluids\/} {\bf 7},
  024301.

\bibitem[Candelier \& Mehlig(2016)]{candelier2016settling}
{\sc Candelier, F. \& Mehlig, B.} 2016 Settling of an asymmetric dumbbell in a
  quiescent fluid. {\em J. Fluid Mech.\/} {\bf 802}, 174--185.

\bibitem[Candelier {\em et~al.\/}(2019)Candelier, Mehlig \&
  Magnaudet]{candelier2019time}
{\sc Candelier, F., Mehlig, B. \& Magnaudet, J.} 2019 Time-dependent lift and
  drag on a rigid body in a viscous steady linear flow. {\em J. Fluid Mech.\/}
  {\bf 864}, 554--595.

\bibitem[Chisholm {\em et~al.\/}(2016)Chisholm, Legendre, Lauga \&
  Khair]{chisholm2016squirmer}
{\sc Chisholm, Nicholas~G., Legendre, Dominique, Lauga, Eric \& Khair,
  Aditya~S.} 2016 A squirmer across {R}eynolds numbers. {\em J.Fluid Mech.\/}
  {\bf 796}, 233--256.

\bibitem[Cox(1965)]{cox1965steady}
{\sc Cox, R.G.} 1965 The steady motion of a particle of arbitrary shape at
  small {R}eynolds numbers. {\em J. Fluid Mech.\/} {\bf 23}, 625--643.

\bibitem[Dabade {\em et~al.\/}(2015)Dabade, Marath \&
  Subramanian]{dabade2015effects}
{\sc Dabade, V., Marath, N.~K. \& Subramanian, G.} 2015 Effects of inertia and
  viscoelasticity on sedimenting anisotropic particles. {\em J. Fluid Mech.\/}
  {\bf 778}, 133--188.

\bibitem[Durham {\em et~al.\/}(2013)Durham, Climent, Barry, Lillo, Boffetta,
  Cencini \& Stocker]{durham2013turbulence}
{\sc Durham, William~M., Climent, Eric, Barry, Michael, Lillo, Filippo~De,
  Boffetta, Guido, Cencini, Massimo \& Stocker, Roman} 2013 Turbulence drives
  microscale patches of motile phytoplankton. {\em Nat. Comm.\/} {\bf 4}, 2148.

\bibitem[Einarsson {\em et~al.\/}(2015)Einarsson, Candelier, Lundell, Angilella
  \& Mehlig]{einarsson2015rotation}
{\sc Einarsson, J., Candelier, F., Lundell, F., Angilella, J.R. \& Mehlig, B.}
  2015 Rotation of a spheroid in a simple shear at small {R}eynolds number.
  {\em Phys. Fluids\/} {\bf 27}, 063301.

\bibitem[Faust \& Gulledge(2002)]{faust2002identifying}
{\sc Faust, Maria~A \& Gulledge, Rose~A} 2002 Identifying harmful marine
  dinoflagellates. {\em Contributions from the United States National
  Herbarium\/} {\bf 42}, 1--144.

\bibitem[Gustavsson {\em et~al.\/}(2016)Gustavsson, Berglund, Jonsson \&
  Mehlig]{gustavsson2015preferential}
{\sc Gustavsson, K., Berglund, F., Jonsson, P.~R. \& Mehlig, B.} 2016
  Preferential sampling and small-scale clustering of gyrotactic microswimmers
  in turbulence. {\em Phys. Rev. Lett.\/} {\bf 116}, 108104.

\bibitem[Gustavsson {\em et~al.\/}(2019)Gustavsson, Sheikh, Lopez, Naso, Pumir
  \& Mehlig]{gustavsson2019theory}
{\sc Gustavsson, K., Sheikh, M.~Z., Lopez, D., Naso, A., Pumir, A. \& Mehlig,
  B.} 2019 Theory for the effect of fluid inertia on the orientation of a small
  spheroid settling in turbulence. {\em New J. Phys.\/} {\bf 21}, 083008.

\bibitem[Hairer {\em et~al.\/}(2000)Hairer, N{\o}rsett \&
  Wanner]{hairer2000solving}
{\sc Hairer, E., N{\o}rsett, S.P. \& Wanner, G.} 2000 {\em Solving Ordinary
  Differential Equations {I} Nonstiff problems\/}, 2nd edn. Berlin: Springer.

\bibitem[Happel \& Brenner(1965)]{happel1981low}
{\sc Happel, J. \& Brenner, H.} 1965 {\em Low {Reynolds} number hydrodynamics:
  with special applications to particulate media\/}. Prentice-Hall.

\bibitem[Hinch(1995)]{hinch1995perturbation}
{\sc Hinch, E.~J.} 1995 {\em Perturbation Methods.\/}. Cambridge University
  Press.

\bibitem[Jiang {\em et~al.\/}(2021)Jiang, Zhao, Andersson, Gustavsson, Pumir \&
  Mehlig]{jiang2021inertial}
{\sc Jiang, F., Zhao, L., Andersson, H.~I., Gustavsson, K., Pumir, A. \&
  Mehlig, B.} 2021 Inertial torque on a small spheroid in a stationary uniform
  flow. {\em Phys. Rev. Fluids\/} {\bf 6}, 024302.

\bibitem[Jonsson(1989)]{jonsson1989vertical}
{\sc Jonsson, P.~R.} 1989 Vertical distribution of planktonic ciliates - an
  experimental analysis of swimming behaviour. {\em Marine~Ecol.~Prog.~Ser.\/}
  {\bf 52}, 39--53.

\bibitem[Kempe \& Fr\"ohlich(2012)]{kempe2012improved}
{\sc Kempe, Tobias \& Fr\"ohlich, Jochen} 2012 An improved immersed boundary
  method with direct forcing for the simulation of particle laden flows. {\em
  Journal of Computational Physics\/} {\bf 231}~(9), 3663--3684.

\bibitem[Kessler(1985)]{kessler1985hydrodynamic}
{\sc Kessler, J.~O.} 1985 Hydrodynamic focusing of motile algal cells. {\em
  Nature\/} {\bf 313}, 218--220.

\bibitem[Kessler(1986)]{kessler1986individual}
{\sc Kessler, John~O} 1986 Individual and collective fluid dynamics of swimming
  cells. {\em Journal of Fluid Mechanics\/} {\bf 173}, 191--205.

\bibitem[Khair \& Chisholm(2014)]{khair2014expansions}
{\sc Khair, A.~S. \& Chisholm, N.~G.} 2014 Expansions at small {R}eynolds
  numbers for the locomotion of a spherical squirmer. {\em Phys. Fluids\/} {\bf
  26}~(1), 011902.

\bibitem[Khayat \& Cox(1989)]{khayat1989inertia}
{\sc Khayat, R.E. \& Cox, R.G.} 1989 Inertia effects on the motion of long
  slender bodies. {\em J. Fluid Mech.\/} {\bf 209}, 435--462.

\bibitem[Kim {\em et~al.\/}(2002)Kim, Baek \& Sung]{kim2002implicit}
{\sc Kim, Kyoungyoun, Baek, Seung-Jin \& Sung, Hyung~Jin} 2002 An implicit
  velocity decoupling procedure for the incompressible {Navier-Stokes}
  equations. {\em International Journal for Numerical Methods in Fluids\/} {\bf
  38}, 125--138.

\bibitem[Kim \& Karrila(2013)]{kim2013microhydrodynamics}
{\sc Kim, Sangtae \& Karrila, Seppo~J} 2013 {\em Microhydrodynamics: principles
  and selected applications\/}. Courier Corporation.

\bibitem[Klett(1995)]{klett1995orientation}
{\sc Klett, J.~D.} 1995 Orientation model for particles in turbulence. {\em
  JAS\/} {\bf 52}, 2276--2285.

\bibitem[Kramel(2017)]{kramel2017nonpsherical}
{\sc Kramel, S.} 2017 Non-spherical particle dynamics in turbulence. PhD
  thesis, Wesleyan University.

\bibitem[Lambert {\em et~al.\/}(2013)Lambert, Picano, Breugem \&
  Brandt]{lambert2013active}
{\sc Lambert, Ruth~A., Picano, Francesco, Breugem, Wim-Paul \& Brandt, Luca}
  2013 Active suspensions in thin films: nutrient uptake and swimmer motion.
  {\em Journal of Fluid Mechanics\/} {\bf 733}, 528--557.

\bibitem[Lauga \& Powers(2009)]{lauga2009hydrodynamics}
{\sc Lauga, E. \& Powers, T.~R.} 2009 The hydrodynamics of swimming
  microorganisms. {\em Rep. Prog. Phys.\/} {\bf 72}, 096601, arXiv: 0812.2887.

\bibitem[Lighthill(1952)]{lighthill1952squirming}
{\sc Lighthill, M.~J.} 1952 On the squirming motion of nearly spherical
  deformable bodies through liquids at very small {R}eynolds numbers. {\em
  Comm. Pure Appl. Math.\/} {\bf 5}, 109--118.

\bibitem[Lopez \& Guazzelli(2017)]{lopez2017inertial}
{\sc Lopez, D. \& Guazzelli, E.} 2017 Inertial effects on fibers settling in a
  vortical flow. {\em Phys. Rev. Fluids\/} {\bf 2}, 024306.

\bibitem[Meibohm {\em et~al.\/}(2016)Meibohm, Candelier, Ros\'en, Einarsson,
  Lundell \& Mehlig]{meibohm2016angular}
{\sc Meibohm, J., Candelier, F., Ros\'en, T., Einarsson, J., Lundell, F. \&
  Mehlig, B.} 2016 Angular velocity of a spheroid log rolling in a simple shear
  at small {Reynolds} number. {\em Phys. Rev. Fluids\/} {\bf 1}.

\bibitem[Menon {\em et~al.\/}(2017)Menon, Roy, Kramel, Voth \&
  Koch]{menon2017theoretical}
{\sc Menon, U., Roy, A., Kramel, S., Voth, G. \& Koch, D.} 2017 Theoretical
  predictions of the orientation distribution of high-aspect-ratio, inertial
  particles settling in isotropic turbulence. {\em Abstract Q36.00011, 70th
  Annual Meeting of the APS Division of Fluid Dynamics, Denver, Colorado\/} .

\bibitem[Pedley \& Kessler(1987)]{pedley1987orientation}
{\sc Pedley, Timothy~John \& Kessler, JO} 1987 The orientation of spheroidal
  microorganisms swimming in a flow field. {\em Proceedings of the Royal
  Society of London. Series B. Biological Sciences\/} {\bf 231}~(1262), 47--70.

\bibitem[Peskin(2002)]{peskin2002immersed}
{\sc Peskin, Charles~S} 2002 The immersed boundary method. {\em Acta
  numerica\/} {\bf 11}, 479--517.

\bibitem[Qiu {\em et~al.\/}(2022)Qiu, Cui, Climent \& Zhao]{qiu2022gyrotactic}
{\sc Qiu, Jingran, Cui, Zhiwen, Climent, Eric \& Zhao, Lihao} 2022 Gyrotactic
  mechanism induced by fluid inertial torque for settling elongated
  microswimmers. {\em Physical Review Research\/} {\bf 4}~(2), 023094.

\bibitem[Redaelli {\em et~al.\/}(2022{\natexlab{{\em a\/}}})Redaelli,
  Candelier, Mehaddi, Eloy \& Mehlig]{redaelli2022hydrodynamic}
{\sc Redaelli, T., Candelier, F., Mehaddi, R., Eloy, C. \& Mehlig, B.}
  2022{\natexlab{{\em a\/}}} Hydrodynamic force on a small squirmer moving with
  a time-dependent velocity at small {R}eynolds number .

\bibitem[Redaelli {\em et~al.\/}(2022{\natexlab{{\em b\/}}})Redaelli,
  Candelier, Mehaddi \& Mehlig]{redaelli2021unsteady}
{\sc Redaelli, T., Candelier, F., Mehaddi, R. \& Mehlig, B.}
  2022{\natexlab{{\em b\/}}} Unsteady and inertial dynamics of a small active
  particle in a fluid. {\em Phys. Rev. Fluids\/} {\bf 7}, 044304.

\bibitem[Roberts(1970)]{roberts1970geotaxis}
{\sc Roberts, A.~M.} 1970 Geotaxis in motile micro-organisms. {\em J. Exp.
  Biology\/} {\bf 53}, 687.

\bibitem[Roberts \& Deacon(2002)]{roberts2002gravitaxis}
{\sc Roberts, A.~M. \& Deacon, F.~M.} 2002 Gravitaxis in motile
  micro-organisms. {\em J. Fluid. Mech.\/} {\bf 452}, 405.

\bibitem[Ros\'{e}n {\em et~al.\/}(2015)Ros\'{e}n, Nordmark, Aidun, Lundell \&
  Mehlig]{rosen2015numerical}
{\sc Ros\'{e}n, T.~Einarsson, J., Nordmark, A., Aidun, C.~K., Lundell, F. \&
  Mehlig, B.} 2015 Numerical analysis of the angular motion of a neutrally
  buoyant spheroid in shear flow at small {R}eynolds numbers. {\em Phys. Rev.
  E\/} {\bf 92}, 063022.

\bibitem[Roy {\em et~al.\/}(2019)Roy, Hamati, Tierney, Koch \&
  Voth]{roy2019inertial}
{\sc Roy, A., Hamati, R.~J., Tierney, L., Koch, D.~L. \& Voth, G.~A.} 2019
  Inertial torques and a symmetry breaking orientational transition in the
  sedimentation of slender fibres. {\em J. Fluid Mech.\/} {\bf 875}, 576.

\bibitem[Schwartz(1966)]{schwartz1966theorie}
{\sc Schwartz, L.} 1966 {\em Theorie des distributions\/}. Paris: Hermann, DL.

\bibitem[Smayda(2010)]{smayda2010adaptations}
{\sc Smayda, TJ} 2010 Adaptations and selection of harmful and other
  dinoflagellate species in upwelling systems 1. morphology and adaptive
  polymorphism. {\em Progress in Oceanography\/} {\bf 85}~(1-2), 53--70.

\bibitem[Subramanian \& Koch(2005)]{subramanian2005inertial}
{\sc Subramanian, G. \& Koch, Donald~L.} 2005 Inertial effects on fibre motion
  in simple shear flow. {\em J. Fluid Mech.\/} {\bf 535}, 383--414.

\bibitem[Titelman \& Ki{\o}rboe(2003)]{titelman2003motility}
{\sc Titelman, Josefin \& Ki{\o}rboe, Thomas} 2003 Motility of copepod nauplii
  and implications for food encounter. {\em Marine Ecology Progress Series\/}
  {\bf 247}, 123--135.

\bibitem[Uhlmann(2005)]{uhlmann2005immersed}
{\sc Uhlmann, Markus} 2005 An immersed boundary method with direct forcing for
  the simulation of particulate flows. {\em Journal of computational physics\/}
  {\bf 209}~(2), 448--476.

\bibitem[{Vesey II} \& Goldenfeld(2007)]{veysey2007simple}
{\sc {Vesey II}, John \& Goldenfeld, Nigel} 2007 Simple viscous flows: from
  boundary layers to the renormalization group. {\em Rev. Mod. Phys.\/} {\bf
  79}, 883--927.

\bibitem[Visser(2011)]{visser2011small}
{\sc Visser, A.} 2011 {\em Small, wet \& rational, individual based zooplankton
  ecology\/}. {DTU Denmark}.

\bibitem[Wang \& Ardekani(2012)]{wang2012inertial}
{\sc Wang, S. \& Ardekani, A.} 2012 Inertial squirmer. {\em Phys. Fluids\/}
  {\bf 24}~(10), 101902.

\end{thebibliography}

\end{document}